\documentclass[12pt]{article}
\usepackage{cite}
\usepackage{graphicx}

\newcommand{\dd}{\mbox{\rm{d}}}
\newcommand{\Tc}{\mbox{$T_{\rm c}$}}

\newcommand{\avp}{\mbox{$\bar{p}$}}

\newcommand{\pr}{\mbox{$p(\vec{r})$}}
\newcommand{\pri}{\mbox{$p(\vec{r_i})$}}

\newcommand{\avpr}{\mbox{$\bar{p}(\vec{r})$}}
\newcommand{\dispr}{\mbox{$p_{\rm disord}(\vec{r})$}}

\newcommand{\deltap}{\mbox{$\Delta p_{\rm disord}$}}
\newcommand{\eps}{\mbox{$\varepsilon$}}

\newcommand{\Ns}{\mbox{$N$}}

\newcommand{\Tcopt}{\mbox{$T_c^{\rm opt}$}}

\newcommand{\TLG}{\mbox{$T_{\rm Gi}$}}
\newcommand{\TKT}{\mbox{$T_{\rm KT}$}}
\newcommand{\AKT}{\mbox{$A_{\rm KT}$}}
\newcommand{\eC}{\mbox{$\varepsilon^c$}}
\newcommand{\TsuperC}{\mbox{$T^c$}}

\newcommand{\LSCOf}{\mbox{La$_{2-x}$Sr$_x$CuO$_{4+y}$}}
\newcommand{\LSCO}{\mbox{LSCO}}
\newcommand{\YBCOf}{\mbox{YBa$_2$Cu$_3$O$_{7-\delta}$}}
\newcommand{\YBCO}{\mbox{YBCO}}
\newcommand{\STOf}{\mbox{(100)SrTiO$_3$}} %<----es asi??

\newcommand{\deltar}{\mbox{$\delta(\vec{r})$}}
\newcommand{\Pmax}{\mbox{$P_{\rm max}$}}
\newcommand{\Pmin}{\mbox{$P_{\rm min}$}}
\newcommand{\RT}{\mbox{$R(T)$}}

\newcommand{\rhon}{\mbox{$\rho_{\rm n}$}}
\newcommand{\rhoninv}{\mbox{$\rho_{\rm n}^{-1}$}}
\newcommand{\Ds}{\mbox{$\Delta\sigma$}}

\newcommand{\Ibias}{\mbox{$I_{\rm bias}$}}
\newcommand{\Ibiassquared}{\mbox{$I_{\rm bias}^2$}}
\newcommand{\Vbias}{\mbox{$V_{\rm bias}$}}
\newcommand{\Vbiassquared}{\mbox{$V_{\rm bias}^2$}}

\newcommand{\Rsq}{\mbox{$R_{\rm sq}$}}

\newcommand{\lsim}{\mbox{$\stackrel{<}{_\sim}$}}

\newcommand{\eg}{{\it e.g.}}
\newcommand{\ie}{{\it i.e.}}
\newcommand{\vs}{{\it vs.\/}}

\newcommand{ \etal}{{\it et al.}}

\newcommand{\Eq}[1]{Eq.\,\ref{#1}}
\newcommand{\Eqs}[1]{Eqs.\,\ref{#1}}
\newcommand{\NoEq}[1]{~\ref{#1}}

\newcommand{\Fig}[1]{Fig.\,\ref{#1}}
\newcommand{\Figs}[1]{Figs.\,\ref{#1}}
\newcommand{\NoFig}[1]{~\ref{#1}}

\newcommand{\Table}[1]{Table~\ref{#1}}

\newcommand{\rmUDonezone}{{UD--1\,zone}}
\newcommand{\rmUDsixzone}{{UD--6\,zone}}
\newcommand{\rmUDtenzone}{{UD--10\,zone}}
\newcommand{\rmUDNzone}{{UD--N\,zone}}
\newcommand{\rmODonezone}{{OD--1\,zone}}
\newcommand{\rmODsixzone}{{OD--6\,zone}}
\newcommand{\rmODtenzone}{{OD--10\,zone}}
\newcommand{\rmODNzone}{{OD--N\,zone}}

\newcommand{\UDonezone}{{\it\rmUDonezone}}
\newcommand{\UDsixzone}{{\it\rmUDsixzone}}
\newcommand{\UDtenzone}{{\it\rmUDtenzone}}
\newcommand{\UDNzone}{{\it\rmUDNzone}}
\newcommand{\ODonezone}{{\it\rmODonezone}}
\newcommand{\ODsixzone}{{\it\rmODsixzone}}
\newcommand{\ODtenzone}{{\it\rmODtenzone}}

\newcommand{\ODNzone}{{\it\rmODNzone}}

\begin{document}
\begin{center}
{\Large\bf Design of structured \LSCOf\  films as superconducting transition-edge  sensors at 4.2K}\\
\end{center}\mbox{}\vspace{-1cm}\\

\begin{center}
\normalsize{M.M. Botana$^1$, A.S. Viz$^2$, M.V. Ramallo$^{1,*}$}
\end{center}

\begin{center}

\normalsize
$^1${QMatterPhotonics, Department of Particle Physics and Institute of Materials iMATUS, University of Santiago de Compostela 15782, Spain.}\\
$^2${QMatterPhotonics,   Department of Particle Physics, University of Santiago de Compostela 15782, Spain.}\\
$^*$Corresponding author (mv.ramallo@usc.es)
\end{center}

\mbox{}\vskip0.5cm\noindent{\bf Abstract:}
We calculate the effects of carrier-density structuration and patterning on thin films of the cuprate  superconductor \LSCOf, in order to optimize its functional characteristics as sensing material for resistive transition-edge bolometers  at liquid-He temperature. 
We perform finite-element computations  considering  two major contributions to  structuration: The intrinsic random nanoscale disorder  associated to carrier density nonstoichiometry, plus  the imposition of regular arrangements of  zones with different nominal carrier densities. Using ad-hoc seek algorithms, we obtain  various structuration designs that  markedly improve the  bolometric performance, mainly  the saturation power and dynamic range. Bolometric operation becomes favorable even in the easier-to-implement constant current mode of measurement. 
\vspace{18pt}

\noindent{\sl\small Graphical Abstract:}\\ \vskip-9pt
\mbox{}\hspace{1cm}\includegraphics[width=5in]{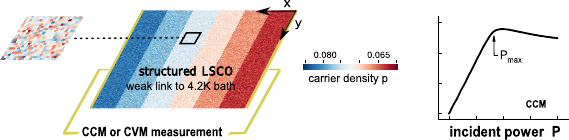}\vskip3pt


\thispagestyle{empty}

\newpage
\setlength{\baselineskip}{18pt}

%%%%%%%%%%%%%%%%%%%%%%%%%%%%%%%%%
%%%%%%%%%%%%%%%%%%%%%%%%%%%%%%%%%
%%%%%%%%%%%%%%%%%%%%%%%%%%%%%%%%%
%%%%%%%%%%%%%%%%%%%%%%%%%%%%%%%%%

\section{Introduction}

Sensing devices based on novel superconducting surfaces\cite{vafapour,yang-intro,amini,brock,badia} such as transition-edge sensors (TES) are  employed in multiple fields of research and technology for detection of radiation, for instance in contexts where broad bandwidth and high sensitivity is primordial and refrigerated operation is feasible or even desirable \cite{lita, ullom, richards, dobbs}. Some examples of application are astrophysics (cosmic background  radiation, etc.)\cite{romani, nagler}, avionics and satellite technology (for environmental imaging, etc.)\cite{harper, jaehnig}, neutron detection for fusion research\cite{appl-1-neutron-detection}, etc. Also the current emergence of quantum technologies of information, often requiring cryogenics to limit quantum decoherence effects, is increasing the interest in  TES  to discriminate  photonic signals.\cite{varnava, charaev, chang} 

Most existing TES  use conventional low-temperature superconductors with very sharp transitions, commonly in thin film or microwire form and operated in constant-voltage mode (CVM) to limit their thermal runaway.\cite{runaway} However, using a cuprate superconductor as sensing material may potentially bring new options of design.\cite{mohajeri,nivelle,probst,neff,neff-II} The probably best-known options for using cuprates  in TES devices involve operating them slightly below its mean-field critical temperature \Tc, often with liquid-N$_2$ cryogenics and sometimes using a non-ohmic regime of their electrical resistivity  and  sensitive amplifiers to achieve readout of the response to incoming radiation.\cite{mohajeri,nivelle,probst,neff,neff-II} 

In a recent paper,\cite{verde} we  made a first exploration of a novel route to further optimize the use of cuprate thin films in TES applications. We considered  custom designs of micro- and nano-scale planar spatial variations of the local carrier concentration \pr, seeking to modify  the profile of the  \RT\  transition  (the electrical resistance versus temperature near the superconducting transition) and thus the corresponding bolometric operational parameters: mainly the temperature coefficient of resistance TCR=$R^{-1}\dd R/\dd T$ and the saturation power \Pmax\ (or rather equivalently its ratio with the minimum detectable power, \ie, the linear dynamic range DR=\Pmax/\Pmin). The cuprates simulated in~\cite{verde} were \YBCOf\ (\YBCO) thin films with different planar  variations \pr\  (that are  experimentally achievable, \eg, through spatial variations of the oxygen index \deltar\ for which various techniques do exist\cite{yang, cordoba,lang-review}). The best obtained  \pr\ designs move the operating temperature to the value 77\,K (convenient for liquid-N$_2$ cryogenics), allow operation in a region with  simple ohmic response readable with conventional electronics  and provide, \eg, about double TCR and one order of magnitude larger DR than  non-structured films.\cite{verde,oppenheim}

However, for some applications  77\,K is not the most desirable operating temperature. Liquid-He temperature 4.2\,K may be  preferred, so to reduce the thermal noise or because the advantageous values of  heat capacity and thermal conductivities of the whole TES setup (that affect aspects such as the recovery time, often depending on temperature-dependent characteristics of the supporting substrates and thermal links rather than on the sensing superconductor itself).\cite{lita,ullom,richards,dobbs,romani, nagler,harper,jaehnig,appl-1-neutron-detection} Therefore, it is interesting to seek for possible \pr\  structuration designs that  optimize cuprates for their use in TES sensors designed for operating temperature 4.2\,K. 

With that aim, in the present paper we explore designs for  micro- and nano-structurations of \pr\ in thin films of the cuprate  \LSCOf\ (\LSCO). In this material, various methods have been experimentally demonstrated to change the carrier concentration $p$, with 4.2\,K well within the range of  attainable critical temperatures.\cite{ando,cooper,mahmood,arpaia,meyer,mosqueira} This includes not only  variation of  the $x$ and/or $y$ stoichiometry indexes during sample growth,\cite{mosqueira} but also using  techniques able of variable local actuation, and hence \pr\ patterning, such as, \eg, focused ion-beam direct writing,\cite{lang,karrer} local long-time SEM exposition,\cite{cho,tolpygo} or surface modification via tip-based scanning microscopies like  STM\cite{yang} or heated-tip AFM\cite{afm}. %, and other\cite{laser}. 

When considering \LSCO, various differences and similarities  with respect to the case of \YBCO\ have to be considered from the beginning. Firstly, the variation of the \RT\  curves with the carrier concentration $p$ are qualitatively different, with \LSCO\ developing for the so-called underdoped $p$-values ($p<0.16$) a smaller and negative TCR at temperatures above the fluctuation transition rounding,\cite{ando} while for overdoped $p>0.16$ it remains positive\cite{ando,cooper,barisic} (in contrast, in \YBCO\cite{arpaia,ando,barisic} the TCR is always positive but the $p>0.16$ range is not  fully accessible by change of the chemical index $\delta$). Also the fluctuation rounding of the \RT\  transition is qualitatively different, as the more pronounced anisotropy and 2D-character of \LSCO\  increase the critical exponents and, notably,  the importance of  the lower tail of the \RT\  transition, due to the Kosterlitz-Thouless phenomenon.\cite{popovic,coton} In addition, the dependence of the critical temperature itself with doping, $\Tc(p)$, has in both compounds a similar general dome-like behavior centered around $p=0.16$ but in the vicinity of $p=0.125$ there is in \LSCO\ a pronounced  local depression, versus a smooth plateau in \YBCO.\cite{timusk,coton-II,shi} (Also heat capacities near \Tc\ are different\cite{loram-II}, mainly due to the temperatures at which they must be evaluated). On the other hand, both materials share the significant importance of disorder effects that, as shown, \eg, in \cite{mosqueira,botana,verde} and will be also confirmed in this paper, will be key in the TES behavior of the structured material.

The outline of this paper is as follows: In Sect.~\ref{Sec:Methods} we describe the methods implemented by us; this includes Subsect.~\ref{Sec:RT} on the modeling of  \RT\ for a given carrier concentration $p$,  Subsect.~\ref{Sec:Pr} on the modelization of local variations \pr\ (composed of both imposed designs \avpr\ plus Gaussian random disorder effects \dispr),  and Subsect.~\ref{Sec:Comput} on our computational methods for seeking optimal \pr\ and calculating their corresponding  bolometric parameters (TCR, DR, etc.). In Sect.~\ref{Sec:OneZone} we describe our results when considering the non-patterned case, both for underdoped  and overdoped \avp-values (and also compare them against some existing $R(T)$ measurements that in this first simple case are available). In Sect.~\ref{Sec:Underdoped} we describe our results for spatial patterns optimizing the bolometric performance for the underdoped case, finding in particular that some spatial pattern designs lead to much better bolometric performance than the non-patterned surfaces  (including, \eg, one order of magnitude larger saturation power and correspondingly larger DR; see \Table{table}). In Sect.~\ref{Sec:Overdoped} we present our corresponding results for  overdoped carrier densities, where again we propose some spatial patterns of much improved bolometric performance (even somewhat better than those of the underdoped case; see also \Table{table}). These results suggest that structured \LSCO\ thin films with ad-hoc planar structurations are promising as sensing surfaces  for TES bolometers at liquid-He temperature. Finally, in our concluding  Sect.~\ref{Sec:Conclusions} we comment on some aspects that we believe could ease the application of our calculations to eventual efforts to experimentally produce such ad-hoc TES sensing surfaces.

%%%%%%%%%%%%%%%%%%%%%%%%%%%%%%%%%%%%%%%%%%%%%%%%%%%%%%%%%%%%%%%%%%%%%%%%%%%%%%%%%%%%%%%%%%%%%%%%%%%%%%%%%%%%%%%%%%%%%%%%%%%%%%%%%%%%%%%%%%%%%%%%%%%%

\section{Methods to find optimal carrier density  structurations of LSCO thin films for better   bolometric sensing performance at $4.2\mbox{K}$\label{Sec:Methods}}

\subsection{Resistive transition of the base non-structured film\label{Sec:RT}}

In order to obtain the electrical resistance transition \RT\ of a \LSCO\ film with a given spatial map of carrier concentrations \pr,  we shall use in this work finite-element computations with each simulation cell having a single local value of $p$. Thus, a needed starting point is the  resistance to be associated to each of those elements. Fortunately, extensive measurements and theory comparisons do exist in the literature  in  \LSCO\ samples of well-controlled and highly homogeneous values  of $p$, that will serve us as valid phenomenological  equations for our purposes. The critical temperature, the normal-state behavior, and the resistivity rounding near the transition are the three aspects  to  be considered  for   such    resistance curves  for each $p$. We describe those aspects below.

%%%
\subsubsection{\label{Sec:Tcp}Critical temperature \Tc\ as a function of carrier density $p$}
It is well-known that the $\Tc(p)$ dependence in \LSCO\ and other cuprates is dome-like, with maximum value reached at $p=0.16$ (the so-called optimal doped $p$-value).\cite{barisic} Also  that this dome is somewhat altered around $p=1/8$,  where a further depression of \Tc\ happens (likely due to stripe order matching\cite{kofu,yamada}). As shown in \cite{coton-II}, for \LSCO\ films this can be accurately accounted for by the empirical formula:
%%%%%%%%%%%%%%
\begin{equation}
\label{Tcp}
\Tc(p) = \Tcopt 
{\Big[} {\big(}1-\frac{\scriptstyle p -0.16}{\scriptstyle 0.11}{\big)}^2{\Big]} 
- \delta T_{c1/8} \exp
{\Big[}-{\big(} \frac{\scriptstyle p-1/8}{\scriptstyle 0.01} {\big)}^2{\Big]}.
\end{equation} 
%%%%%%%%%%%%%%%
Regarding the values of \Tcopt\ and  $\delta T_{c1/8}$ let us note already here that in this paper we shall use material parameters corresponding to \LSCO\ films of thickness  100\,nm grown over  \STOf\ substrates, a typical morphology appropriate for TES applications. This corresponds to $\Tcopt=23$K and $ \delta T_{c1/8}=4.8$K.\cite{tlmeyer,coton-II}  (We also note already here that we shall defer to  Sect.~\ref{Sec:Conclusions} on a discussion on the parameter values and other considerations that should be fine-tuned in our numerical analyses to extend them to \LSCO\ films not matching exactly our default parameter choices, like film thickness, etc. Moderate quantitative, but not qualitative,  differences are obtained in our final results when considering changes in such assumed details.)

%%%
\subsubsection{Normal-state contribution to the electrical resistivity}
It is customary\cite{popovic,coton,timusk,rey,mgbdos,isotropic,klemm,puica,carballeira} to describe the electrical resistivity $\rho$ of superconductors at temperatures above and near \Tc\ as two separate additive contributions to its reciprocal, the electrical conductivity, in particular as $\rho^{-1}=\rhoninv+\Ds$ where  \rhon\  is the normal-state resistivity, that may be obtained from measurements well above \Tc,  and \Ds\ is the so-called paraconductivity, or fluctuation conductivity, producing a rounding near \Tc\ of the $\rho(T)$ transition.

Regarding \rhon, in cuprate superconductors its phenomenology as a function of $T$ and $p$ is today quite comprehensibly described at the empirical experimental level\cite{ando,cooper,arpaia,timusk} (even if the reason for such phenomenology is famously not yet truly understood, being a problem almost as open as the origin of the high \Tc\ itself\cite{hussey,timusk}). Following standard practice,\cite{cooper,ando,coton,timusk} will use the following formulas:
\begin{eqnarray}{lll}
\label{bkguno}
\rhon(T,p\leq0.15) &=& A(p) + B(p) T + C(p) T^2 + D(p)/T,\;\; \\
\rhon(T,p>0.15) &=& A(p) + B(p) T + C(p) T^2.
\label{bkgdos}
\end{eqnarray}
For the coefficients $A(p)$, $B(p)$, $C(p)$ and $D(p)$ we will use the numerical values that result from fitting the extensive set of experimental measurements in \LSCO\ films reported  for $p\leq0.22$ by Ando  \etal\cite{ando} and for $p>0.23$ by Cooper  \etal\cite{cooper}, and interpolate them for the intermediate $p$ values. We preformed these fits in the  region $30\,{\rm K} \lsim T \lsim 150\,{\rm K}$ (because for lower temperatures the resistivity data may be affected by the rounding due to critical fluctuations near the transition, as we discuss in the next subsection). 

\subsubsection{Paraconductivity contribution}
Regarding the \Ds\ contribution to $\rho^{-1}$, for cuprates again ample literature exist (see, \eg, \cite{coton,botana,klemm,puica}) proposing equations that phenomenologically describe it with excellent accuracy (irrespectively of possible differences in the theory interpretations, or even the existence of alternative formulas possibly also able to accurately describe these data\cite{rey}). We will employ in this paper the same set of equations as we used in \cite{coton} for previous studies of \Ds\ in LSCO films as a function of doping, covering the whole range of temperatures where the rounding of the transition is observed, including: the lower tail of the transition,  associable\cite{barisic,popovic,coton} with a Kosterlitz-Thouless (KT) regime of the fluctuations; the main transition region around the $\rho(T)$ inflection point,  associable\cite{puica,klemm,rey} with the Gaussian regime of the fluctuations; and the high-temperature part of the rounding, extending up to about 2\Tc\ and describable\cite{carballeira} in terms of short-wavelength, or high-energy, fluctuations. As shown in \cite{coton}, the following equations are in excellent agreement with observations in those regions: 
%%%%%%%%%%%%%%%
\begin{eqnarray}{lcl}
\label{BKT}
\Delta \sigma = \AKT\,  \exp\sqrt{ 4  \frac{\Tc -\TKT}{T -\TKT}} &&\mbox{\rm\; for } \TKT<T<\TLG,\\
\Delta \sigma = \frac{e^2}{16\hbar d} \frac{1}{\eps}\Big( 1 - \frac{\eps}{\eC} \Big)^2 &&\mbox{\rm\; for } \TLG\leq T\leq\TsuperC.
\label{paraGGL}
\end{eqnarray}
%%%%%%%%%%%
The parameters involved in these equations, and numerical values for them  in agreement\cite{coton} with data in \LSCO\ films, are as follows: \Tc\ is the mean-field critical temperature (given by \Eq{Tcp} in \LSCO\ films);  \TKT\ is the Kosterlitz-Thouless temperature (located in the tail of the transition, found in \LSCO\ films to be well approximated by  \Tc-2\/K\cite{coton});  \TLG\ is the Ginzburg temperature (the boundary above \Tc\ between the KT and Gaussian fluctuation regimes, $\TLG=1.015\, \Tc$ in \LSCO\  films\cite{coton,loram});  \TsuperC\ is the cutoff temperature well above the transition where fluctuation effects become negligible, that we may take as $\TsuperC=1.7\Tc$;\footnote{Actually, in \cite{coton} it is found a larger value for \TsuperC\ in the underdoped case, but $\TsuperC=1.7\Tc$  fits well the data also in that case when restricting the analyses to $T<30\,{\rm K}$, as in our present paper.}\cite{coton,rey,mgbdos,isotropic}    $\eps=\ln(T/\Tc)$ is a reduced temperature;   $\eC=\ln(\TsuperC/\Tc)$;   $d$ is the distance between superconducting layers ($6.6$\,\AA\  in \LSCO\cite{mosqueira,coton-II})   and \AKT\ is a proportionality constant that may be obtained by equating \Eqs{BKT} and\NoEq{paraGGL} at $T=\TLG$. \cite{coton,botana}

\subsection{Consideration of spatial maps of local carrier density: nominal pattern and disorder contributions\label{Sec:Pr}}

The main subject of study of the present work is  \LSCO\ films with a spatial variation \pr\ of carrier concentrations, instead of the homogeneous $p$ value assumed above. We shall consider in what follows two sources of \pr\ variation, namely:
%%%%%%%%%%%%%
\begin{equation}
\label{pcontributions}
\pr=\avpr+\dispr,
\end{equation}
%%%%%%%%%%%% 
where \avpr\ is the nominal carrier concentration map, \ie, the ad-hoc pattern externally imposed by the experimentalist using any structuration procedure of choice (\eg, focused ion-beam direct writing, local SEM oxidation, etc.) and \dispr\ is a spatially random  contribution related to the unavoidable disorder in any actual \LSCO\ sample (a pictorial view of this two-scale structuration  is given in our graphical abstract).

The main objective of the present paper is to research the \avpr\ patterns optimizing the bolometric performance. Note, however,  that considering  the \dispr\ contribution is also crucial  for an accurate description of the resistivity in the transition region. This has been  shown to be the case even for the best homogeneous \LSCO\ films grown by different experimentalists (see, \eg, Fig.~7 of \cite{coton-II}). The reason for the unavoidable appearance of the \dispr\ contributions even for non-patterned (\ie, \avpr=const) films may be traced to the non-stoichiometry of the  carrier concentrations (in the superconducting range $0.05\lsim p \lsim 0.27$, where as always we express $p$-values normalized as number of carriers per CuO$_2$ unit cell) that in turn implies that the spatial distribution of carrier donors is nonuniform at the few-unit-cells scale. The statistics of such minimum (or intrinsic) disorder may be calculated using  coarse-grained averages (see \cite{mosqueira,coton-II}), resulting for \LSCO\ in a Gaussian distribution with full-width at half-maximum
%%%%%%%%%%%%%%
\begin{equation}
\label{dp}
\deltap = 0.0162 \sqrt{\bar{p}-\bar{p}^2/2}.
\end{equation}
%%%%%%%%%%%%
The number prefactor results from considering a coarse-grain average size of $(30\, \mbox{nm})^2$, in agreement with direct measurements\cite{mihailovic} of the spatial charge distribution in cuprates.\footnote{It is easy to translate the $p$ distribution into a \Tc\ one, of which magnetic susceptibility measurements allow direct experimental probe. As shown in \cite{mosqueira,coton-II}, such measurements in LSCO samples by different authors do present \Tc\ distributions in accordance with \Eq{dp}.}

\subsection{Numerical computation of  \RT\ of the structured superconducting films and algorithmic search of structurations optimized for bolometric performance\label{Sec:Comput}}

Let us  describe  our computational methods to calculate for each structured film its \RT\ curve, and also our algorithms to search the  \avpr\ patterns that best optimize the bolometric performance when the film is used as a TES sensor element.

To calculate  \RT\ we perform a  finite-element modeling of the  film, in which we assign to each spatial node $i$ of the simulation  a single $p_i$ value and a corresponding  $R_i(T)$. For the \pri\ map we shall consider both the nominal and disorder contributions, as in \Eq{pcontributions}. We then used in-house software  to calculate, using a $200\times200$ finite-element mesh-current matrix method, the global resistivity of the entire nonhomogeneous film. This involves, for each temperature and pattern, inverting matrices of about $10^4$ non-null elements. This task can be parallelized for which we employ a custom cluster (16 GPU cards with more than 2500 floating-point calculators each, allocated to this calculation exclusively, so the computing effort may be maintained over several weeks).  An automated script automatically generates different nominal charge concentration patterns, adds the disorder contributions, launches their $R(T)$  simulations, and selects the nominal pattern maps that yield the best performance for bolometric sensing. 

In the case of \LSCO, as mentioned above, each \Tc\ mono\-do\-main is estimated  to be of area $(30\,\mbox{nm})^2$.\cite{mosqueira,mihailovic,coton-II}. Then our  $200\times200$ finite-element mesh corresponds to  $(6\,\mu\mbox{m})^2$,  an area similar to the  microsensor TES devices implemented by several authors.\cite{niemack,nivelle,probst,miller,sergeev}. We checked in a random selection  of our simulations that increasing the underlying mesh to $400\times400$  elements strongly increments the computation time but leads to the same results.

We  tried different algorithms to generate the different nominal charge concentration patterns \avpr. We obtained our best results from one of the simplest options, consisting in considering samples composed of a finite number \Ns\ of successive slices, transversal to the current's direction, each with carrier concentration $\avp_j$ following a general power law in the slice number $j$, and commanding the algorithm to scan values for the parameters $\avp_{\rm min}$, $\avp_{\rm max}$, \Ns, $\beta$ and $\gamma$ for the power law when written as:
%%%%%%%%%%%%
\begin{equation}
\label{pj}
\avp_j = \avp_{\rm min} + \left[\frac{ \avp_{\rm max}- \avp_{\rm min}}{\left(N-1\right)^\beta}\right] j^\beta, \;\;\;\;\;\; j=0,\dots\Ns-1,
\end{equation}
%%%%%%%%%%%%%%%%%
 where each successive slice is of thickness $x_{j+1}-x_j$ and
 %%%%%%%%%%%%%%%
\begin{equation}
\label{lambdap}
x_j = \Gamma \cdot (\avp_j - \avp_{\rm min})^{\gamma},  \;\;\;\;\;\;\;\; j=0,\dots\Ns.
\end{equation}
%%%%%%%%%
(The normalization $\Gamma$ is  given by the film's length). In a second pass of optimization, the algorithm adds to \Eqs{pj} and\NoEq{lambdap} small fine-tuning corrections of a few percent, to further improve the best results identified in the first pass. The scanning formulas\NoEq{pj} and\NoEq{lambdap} are inspired by the results of our previous work for the superconductor \YBCO,\cite{verde} as they have the parameter freedom to reproduce almost exactly the patterns found in that work to optimize the bolometric performance of that material (plus additional  rather different shapes). Note that in the above formula the dependence of \avpr\ is only in the axial coordinate $x$; however, the additional disorder contribution \dispr\ is 2D (again this is the same as in \cite{verde}; note also that this 2D character increases the order of magnitude of the computational effort but, as mentioned before,  is crucial to obtain accurate results in cuprates).

%%%%%%%%%%%%%%
 \begin{table*}\scriptsize
 \caption{\footnotesize Best obtained structurations for improved TES sensor performance.}
\centering
\mbox{}\\
%\begin{tabular}{|l|r|l r|r|r||c|rr|rr|rr|}
\begin{tabular}{l|rl rrrcrrrrrr}

  &   &    &  &  &  & TCR & \multicolumn{2}{c}{\Pmin\ (pW)} & \multicolumn{2}{c}{\Pmax\ ($\mu$W)}& \multicolumn{2}{c}{DR}\\ 
 structuration  & \Ns  &  $\avp_{\rm min}$  & $\avp_{\rm max}$  &  $\beta$\;\mbox{} & $\gamma$\;\mbox{}   &  (K$^{-1}$) & CCM & CVM & CCM & CVM & CCM & CVM  \\
\\
 
\rmUDonezone &  1   &   \multicolumn{2}{c}{0.060}  & --     &  --      & 2.2 & 0.47 & 9.5  & 10  & 5.0 & 2$\,\cdot10^4$ & 5$\,\cdot10^2$\\
\rmUDsixzone &  6   &   0.060   &  0.082  &  1.0  &  1.23   & 3.5   & 0.29  & 28   & 83  & 55  & 3$\,\cdot10^5$ & 2$\,\cdot10^3$\\
\rmUDtenzone & 10   &   0.060   & 0.109 &  1.0    &  1.23   & 3.0    & 0.33  & 38   & 140 & 100 & 4$\,\cdot10^5$  & 3$\,\cdot10^3$ \\

\rmODonezone&  1   & \multicolumn{2}{c}{0.260}  &  --   &  --    & 2.7 & 0.38  & 11  & 15  & 10  & 4$\,\cdot10^4$  & 9$\,\cdot10^2$\\
\rmODsixzone&  6   &   0.231   &  0.255  &  0.80  &  1.55       & 3.5 & 0.20  & 8.0   & 87  & 50  & 4$\,\cdot10^5$  & 6$\,\cdot10^3$\\
\rmODtenzone& 10   &   0.160   & 0.255  &  0.57  &  3.40      & 3.5 & 0.28  & 11   & 190 & 60  & 7$\,\cdot10^5$  & 6$\,\cdot10^3$   \\ 
\end{tabular}
\begin{tabular}{c}
\multicolumn{1}{p{\textwidth}}{\vskip3pt\footnotesize The first group of columns indicates the parameters describing (via \Eqs{pj} and\NoEq{lambdap}) the best structurations found by our algorithms and  simulations for the   underdoped (UD) and overdoped (OD) case and for different number of pattern  zones $N$. The second group summarizes the obtained bolometric  parameters (see also \Figs{fig:UD} and\NoFig{fig:OD})   for  both the constant current  (CCM) and constant voltage  (CVM) modes of operation. Note the  increase in  dynamic range DR achieved by progressive structuration, associated mainly to a increase of \Pmax\ (without significant penalty in  TCR and \Pmin, that sometimes even improve modestly).\vskip3pt\mbox{} }
\end{tabular}
\label{table}
\end{table*}
%%%%%%%%%%%%%%

To obtain the bolometric performance for each simulated film from its computed $R(T)$ we calculate its response to an incoming radiation power. For that, we will use a common thermal model for  TES devices (see,  \eg,  \cite{appel,neff,kraus}) to calculate their stationary state temperature $T_s$ under such radiation:
%%%%%%%%%%%%
\begin{equation}
\label{heat-flow}
P  + P_{\rm self}(T_s) = G \cdot (T_s - T_{\rm bath}),
\end{equation}
%%%%%%%%%%%
where $P$ is the incoming radiation power, $P_{\rm self}$ is the self-generated power by Joule effect,  $T_{\rm bath}$ is the bath temperature and $G$ the thermal conductance of the link to that bath. In accordance with our focus in this paper  we take $T_{\rm bath}=4.2$\,K, and a  value $G = 10^{-5}$W/K  typical\cite{neff-II} for bolometers at that temperature. (We are also making in \Eq{heat-flow}   the common approximation of an optical absorptivity $\eta=1$; otherwise $P$ in our results should be interpreted as $\eta P$\cite{kraus}).  For  $P_{\rm self}$, we will perform our calculations in each of two  possible  operational modes of the sensor:  constant-current mode (CCM) and  constant-voltage mode (CVM). In them,\cite{neff}
%%%%%%%%%%%%
\begin{eqnarray}{ccr}
P_{\rm self} (T)&=& \Ibiassquared\,R(T) \hspace{1.7em}\mbox{\rm for CCM},\label{PselfCCM}\\
P_{\rm self} (T)&=& \Vbiassquared/R(T) \hspace{1.2em}\mbox{\rm for CVM}\label{PselfCVM},
\end{eqnarray}
%%%%%%%%%%%
where \Ibias\ and \Vbias\ are the bias current and voltage imposed in each mode. The CVM operation is often favored for TES devices that employ low-temperature superconductors, to moderate  their problematic  thermal  run\-aways.\cite{runaway} In high-\Tc\ cuprates these are not so much of an issue.\cite{charaev,chang,mohajeri,neff} The CCM  is normally easier  to implement than CVM (that often requires implementing very-low noise reading equipments such as SQUID devices\cite{fukuda}). For completeness, we shall study our proposed material optimizations  in both CVM and CCM.

\begin{figure*}
\centering
  \includegraphics[width=0.95\linewidth]{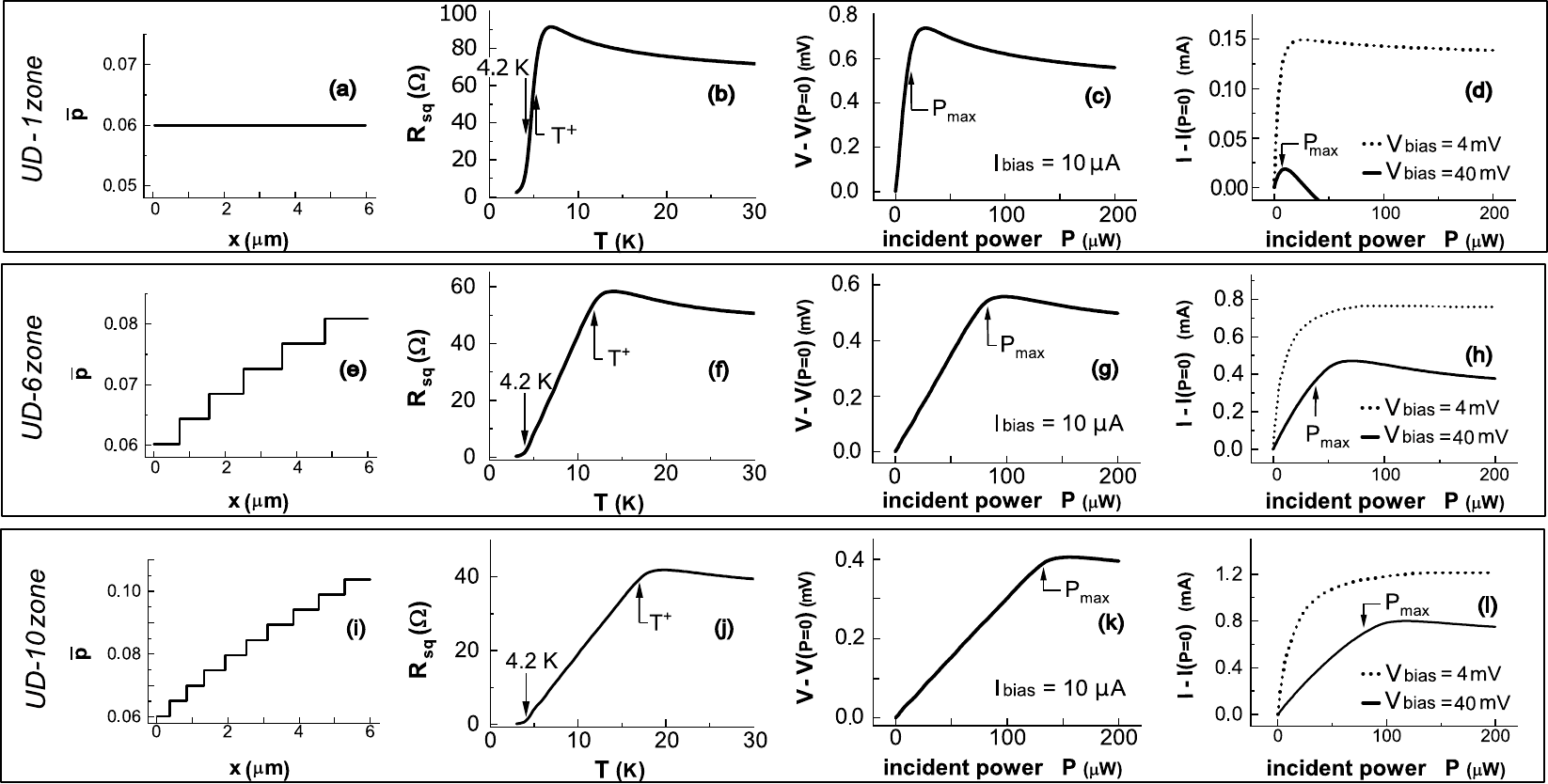}
\caption{\footnotesize Results for sensing films with nominal carrier  densities in the underdoped (UD) range $\avp<0.16$. First row corresponds to a non-patterned film ($N=1$). Subsequent rows correspond to films patterned with $N=6$ and~$10$ zones of \avp \ (following \Eqs{pj} and\NoEq{lambdap} with parameter values of \Table{table}). First column of graphs display \avp\ \vs\ the longitudinal position, second column the electrical sheet resistance \vs\ temperature, third column the output voltage for fixed current (CVM operation mode) \vs\ incident radiation power, and fourth column the output current for constant voltage mode (CVM). $T^+$ and \Pmax\ mark the end of the linear response range. The bias 40mV (solid line in panels d, h and l) optimizes \Pmax\ in CVM operation.}\label{fig:UD}
\end{figure*}

Combination of \Eqs{heat-flow} to\NoEq{PselfCVM} with the $R(T)$ curve allows to calculate the output signal (intensity in CVM and voltage in CCM) produced by the sensor  as a function of the incident optical power $P$ (see  \Figs{fig:UD} and \NoFig{fig:OD}). 

From those results we  obtain  the quantities that, as noted in our Introduction, are the main  parameters usually employed to discuss TES designs: \Pmin, \Pmax, DR and TCR (see \Table{table}).  

In particular, the temperature coefficient of resistance TCR for an incoming power $P$ is given by $\mbox{TCR}=R^{-1}{\mbox{\footnotesize P=0}}\, \delta R/ \delta T$ with $\delta R=R(P)-R{\mbox{\footnotesize P=0}}$ and $\delta T=T(P)-T{\mbox{\footnotesize P=0}}$. As customary,\cite{kraus,neff-II,clarke} we will pursue a constant TCR value for all the operating range of  incident power, what corresponds in practice to a linear dependence of $R$ with $T$ in that range. Therefore, we compute such single TCR value  from  the increments in the whole linear region:
%%%%%
\begin{equation}
\label{TCR}
\mbox{TCR} = \frac{1}{R(T_{\rm bath})} \frac{R(T^+) - R(T_{\rm bath})}{T^+ - T_{\rm bath}},
\end{equation}
%%%%%%
where $T^+$ is the highest temperature up to which  $R(T)$  remains linear in $T$ (see second column of \Figs{fig:UD} and\NoFig{fig:OD}).

The minimum and maximum incident radiation powers that the TES is able to sense, \Pmin\ and \Pmax, are obtained as follows: First we calculate, using \Eqs{Tcp} to\NoEq{PselfCVM}, the measurable output (intensity in CVM and voltage in CCM)  of the sensor to each $P$ value (third and fourth columns of \Figs{fig:UD} and\NoFig{fig:OD}). Then we identify \Pmax\ as the maximum power up to which that measurable output remains linear in $P$ (in our results we will find that incoming powers beyond \Pmax\ not only break the linearity of the sensor but also shortly saturate the measured output anyway; therefore, we did not consider the complication of extending the operation range through nonlinear behaviors).

To estimate the minimum detectable power, \Pmin, in rigor the resolution and sensitivity of the  equipment used to probe the electrical resistivity of the superconductor  must be taken into account, and not only the response of the sensing material. Of course, much can be said about the state-of-art techniques to optimize that reading measurement\cite{richards,fukuda,kraus,neff-II,clarke} and in principle nothing prevents deploying them over the structured films studied here. However, for simplicity and ease of comparison we decided to fix, in our calculations, a modest relative measurement resolution of $10^{-4}$ for both current and voltage signals, as easily attainable with conventional equipment.\cite{fukuda,clarke} Using this common fixed reference will allow us to compare the performances of the different material structurations proposed in this paper with each other. For an example of experimental TES  implementation with a similar resolution see, \eg, \cite{clarke}. 

Finally, the dynamic range is obtained as
$\mbox{DR} = {P_{\rm max}}/{P_{\rm min}}$.

%%%%%%%%%%%%%%%%
%%%%%%%%%%%%%%

%%%%%%%%%%%%%%%%%%%%%%%%%%%%%%%%
%%%%%%%%%%%%%%%%%%%%%%%%%%%%%%%%
%%%%%%%%%%%%%%%%%%%%%%%%%%%%%%%%
\section{Results for non-patterned superconductors (\UDonezone, \ODonezone)\label{Sec:OneZone}}
%%%%%%%%%%%%%%%%%%%%%%%%%%%%%%%%
%%%%%%%%%%%%%%%%%%%%%%%%%%%%%%%%
%%%%%%%%%%%%%%%%%%%%%%%%%%%%%%%%
Let us first present  our results for  the bolometric performance of \LSCO\ thin films without any imposed patterning (\ie, \avpr=const), that are obviously the easiest to fabricate from the experimentalists' point of view. As in the rest of this paper, we seek operational temperatures of 4.2\,K. This corresponds to either $\avp=0.06$ (underdoped case) or $\avp=0.26$ (overdoped). We label these cases in our figures and tables as \UDonezone\ and \ODonezone\ respectively (signaling the under/overdoped case and the number of planar regions with different \avp). 

The results of our simulations for the \UDonezone\ case are shown in the first row of \Fig{fig:UD}. In particular, \Fig{fig:UD}(b) displays the computed sheet resistance $\Rsq(T)$ curve. We mark  with arrows in that figure the bath temperature 4.2\,K and the highest temperature $T^+$ up to which $\Rsq(T)$ is linear in the transition, used to calculate TCR in \Eq{TCR}. In \Fig{fig:UD}(c) we show the  output voltage of the TES sensor as a function of incident optical power $P$, assuming CCM  operation with a typical value of the bias intensity,  $\Ibias=10\mu{\rm A}$. We indicate in that figure   the \Pmax\ obtained from that curve. In \Fig{fig:UD}(d) we assume  CVM operation and show the corresponding output intensity as a function of $P$. When plotting the CVM outputs  in this paper we will always display the results for two different bias \Vbias: One that maximizes the obtained \Pmax\ and a second one that illustrates that using different \Vbias\ would change the shape of the output curve and thus degrade (to a different degree depending on the considered film) the range of incident power for which linear and non-saturated response is observed (a feature that is not observed for the CCM mode).\footnote{The reason why a variation of \Vbias\ affects \Pmax\ is: For a bias that is too low the output $I$ grows in CVM mainly  as $R^{-1}$ and hence is not linear in $P$, while for a bias that is too large the range of significant response to incident power is reduced because $T{\mbox{\scriptsize P=0}}$ is already well above 4.2\,K due to self-heating in \Eqs{heat-flow} to\NoEq{PselfCVM}. Thus, the optimum \Pmax\ is achieved for an intermediate value of \Vbias. (For CCM instead, our \Ibias\ values  produce small self-heating and the output grows with $R$).}

The numerical values resulting from those simulations for the main bolometric performance parameters are summarized in \Table{table}, including TCR, \Pmin, \Pmax\ and  dynamic range DR for both CCM and CVM operations. As may be seen in those results, the non-patterned film with underdoped carrier density $\avp=0.060$ is valid as a bolometer with operating temperature 4.2\,K, but with a  modest performance (ultimately associated with its low value of $T^+$, of about just 1\,K above 4.2\,K).

\begin{figure*}
\centering
  \includegraphics[width=0.95\linewidth]{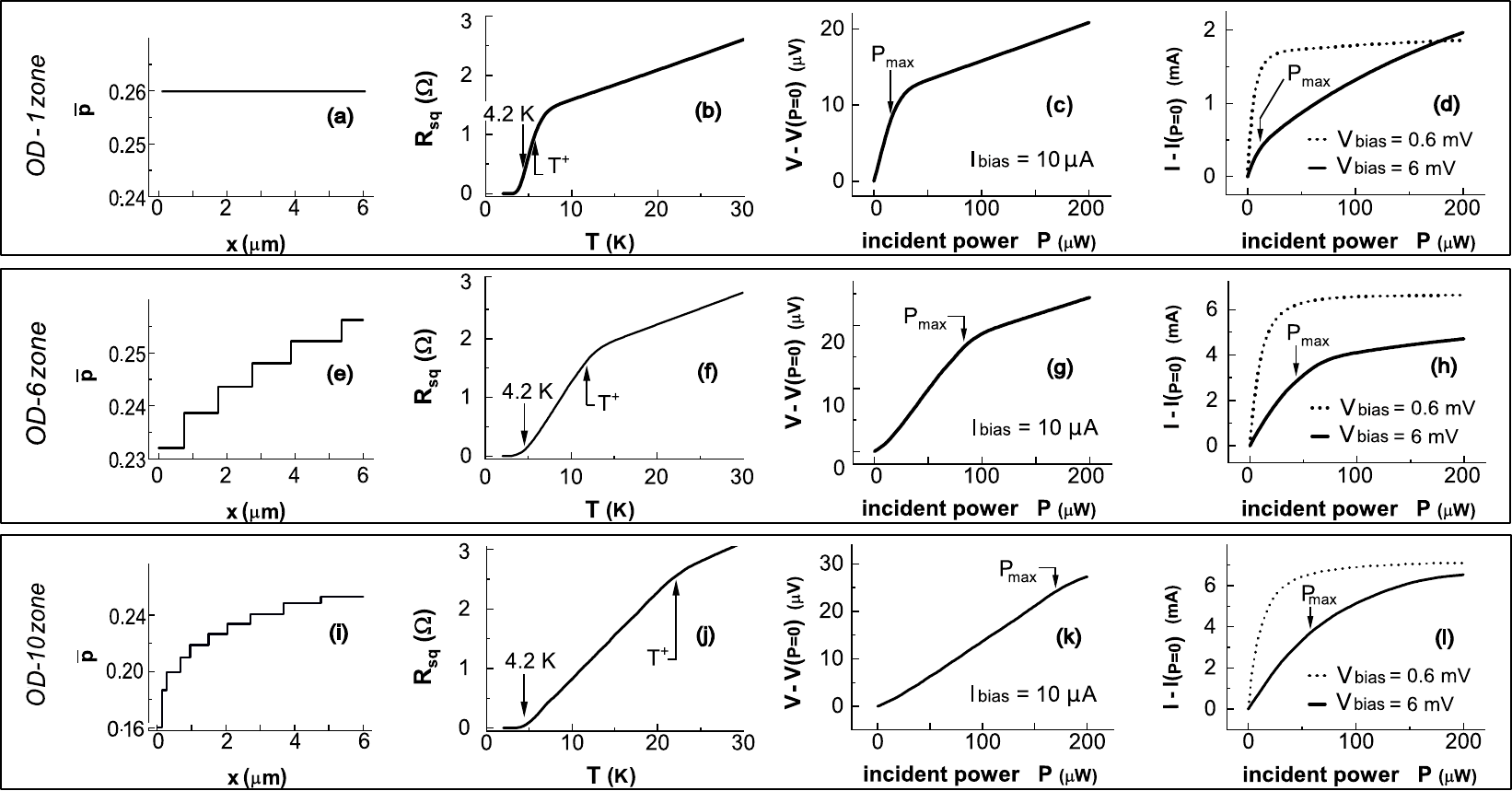}
\caption{\footnotesize Results for films with nominal carrier  densities in the underdoped (OD) range, $\avp>0.16$. Symbol meanings and panel organization are the same as in \Fig{fig:UD}. The bias 6mV (solid line in panels d, h and l) optimizes \Pmax\ in CVM operation.}
\label{fig:OD}
\end{figure*}

Our similar results for the non-patterned film with overdoped carrier density $\avp=0.260$, labeled \ODonezone, are shown in the first row of \Fig{fig:OD} and \Table{table}. Again the low  $T^+$ induces a quite modest bolometric performance, much worse than the ones we  shall obtain for planar-patterned \avpr\ films. 

An interesting difference between the underdoped and overdoped cases is that \Pmax\ is achieved at quite different \Vbias\  when in CVM mode (see  \Figs{fig:UD}(d) and\NoFig{fig:OD}(d)). The reason may be traced back to their  normal-state background contributions to the electrical resistivity (different to each other  in magnitude and $T$-behavior, compare \Figs{fig:UD}(b) and\NoFig{fig:OD}(b)).

\begin{figure}[bh]
\centering
\includegraphics[width=0.35\linewidth]{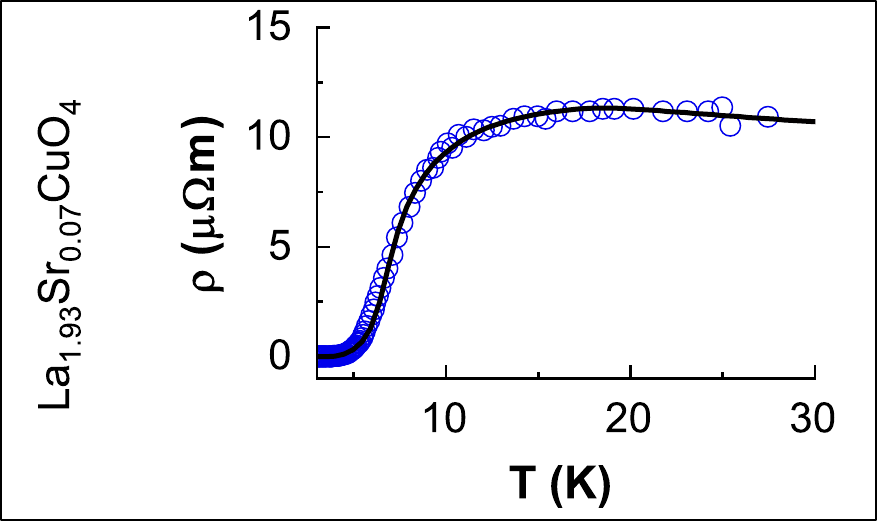}
\caption{\footnotesize Comparison of the electrical resistivity (circles) measured by X.~Shi and coauthors (taken from \cite{shi}) in a \LSCO\ films with \Tc\ of about 4.2\,K, with our model equations  (solid line). The Sr content estimated in~\cite{shi} for this sample was about 0.07 and we employed carrier densities within about 10\% of that value so to better match  \Tc\  (as the exact  value of the latter  depends also on the precise film thickness and chemistry of the  substrate supporting the film, see Sect.~\ref{Sec:Conclusions}).  }
\label{fig:test}
\end{figure}

Finally for the non-patterned films, let us note here that they give us the chance to test the validity of our $R(T)$ simulation procedures against experimental results. This is because measurements of $R(T)$ do exist in the literature for non-patterned \LSCO\ films with \Tc\ near the value 4.2\,K. In \Fig{fig:test}  we compare the experimental data measured by Shi and coauthors (taken from~\cite{shi}) and our simulation results done with the methods described before and assuming a \avp\ value within about 10\% of the one estimated in~\cite{shi} for that sample. The agreement is excellent.  Unfortunately,  no direct   measurements of \Pmax\ and \Pmin\ seem to exist  in a film of that compostion, to our knowledge never yet experimentally implemented as  sensor element in a TES device. For other \avp\  values (but then  \Tc\ quite different from 4.2\,K) the interested reader may find in Ref.~\cite{coton}  further comparisons between experimental $R(T)$ data in non-patterned films and computer simulations using a methodology not too different from the one in the present work, further confirming the basic validity of our present simulation procedures.

%%%%%%%%%%%%%%%%
%%%%%%%%%%%%%%

%%%%%%%%%%%%%%%%%%%%%%%%%%%%%%%%
%%%%%%%%%%%%%%%%%%%%%%%%%%%%%%%%
%%%%%%%%%%%%%%%%%%%%%%%%%%%%%%%%
\section{Results for patterned underdoped superconductors (\UDNzone)\label{Sec:Underdoped}}
%%%%%%%%%%%%%%%%%%%%%%%%%%%%%%%%
%%%%%%%%%%%%%%%%%%%%%%%%%%%%%%%%
%%%%%%%%%%%%%%%%%%%%%%%%%%%%%%%%

We move now to the case in which a \avpr\ spatial pattern is imposed, considering in this Section the underdoped range $\avp<0.16$. 

As discussed in Sect.~\ref{Sec:Methods}, we run simulations that scan numerous different patterns, and select the ones with better bolometric performance. In doing so, we found that a pattern consisting of a relatively modest number  $\Ns=6$ of spatial zones of different nominal carrier concentrations is already able to provide significantly better results than non-patterned films. We label that six-zone   structure  as \UDsixzone. The  main parameter values in \Eqs{pj} and\NoEq{lambdap} defining the patterning are listed in \Table{table}. Also, the $\avp(x)$ profile is plotted  in \Fig{fig:UD}(e) (and it is the example pictured in our graphical abstract)

The bolometric behavior for such structured film is shown in the second row of \Fig{fig:UD}. In particular,   \Fig{fig:UD}(f) displays the  computed $\Rsq(T)$ curve, \Fig{fig:UD}(g) the resulting output versus incident power for CCM  operation and \Fig{fig:UD}(h) the  corresponding output versus incident power for CVM. The resulting performance parameters are summarized in   \Table{table}.

It is easy to see that the obtained sensing characteristics  display a very significant improvement when compared with the non-patterned  \UDonezone: Notably, the dynamic range improves by about one order of magnitude in both  CCM and CVM operation, mainly driven by the significant increase of \Pmax\ in both modes (\Pmin, on the other hand, experiences more moderate changes). The temperature range of operation is also significantly  widened ($T^+$ becomes now about 8\,K above liquid-He temperature). Note also  that the increase in DR does not come at the expense of weaker TCR: On the contrary, \Table{table} shows that TCR is in fact also improved (by more than 50\%).

We also report (see \Fig{fig:UD}(i) to~(l) and \Table{table}) on the results obtained for a probably more difficult-to-fabricate pattern with  $\Ns=10$ zones (labeled as \UDtenzone) that provides even better results. This pattern further extends the linear operation range to be about $T^+-4.2\,{\rm K}\simeq14\,{\rm K}$ and also further  improves \Pmax, without dramatically affecting \Pmin\ or  TCR. The dynamic range DR is correspondingly  enlarged over the one of the simpler 6-zone structuration. That being said, the improvements of \UDtenzone\ vs.\,\UDsixzone\ are well smaller than the ones of \UDsixzone\ vs.\,\UDonezone.

Very noticeable is the fact that the above quantities for DR are very competitive even when compared with the best values  being reported in the literature for TES bolometers  for 4.2\,K operation using low-\Tc\ superconductors (see, \eg, \cite{fukuda}).  Note also that we obtain those good performances  already in CCM operation. In fact, they are better for CCM than for CVM. As already commented, CVM is usually more difficult to effectively implement. Our results suggest that operating a planar-patterned cuprate superconducting thin film in its ohmic linear-in-$T$ region with a simple reading based on good quality but conventional volt-meters of about $10^{-4}$ relative resolution may provide an alternative path for TES bolometer designs at liquid-He base operation temperature.
%
%
%%%%%%%%%%%%%%
%%%%%%%%%%%%%%
%
%
%
%
%%%%%%%%%%%%%%%%%%%%%%%%%%%%%%%%
%%%%%%%%%%%%%%%%%%%%%%%%%%%%%%%%
%%%%%%%%%%%%%%%%%%%%%%%%%%%%%%%%
\section{Results for patterned overdoped superconductors  (\ODNzone)\label{Sec:Overdoped}}
%%%%%%%%%%%%%%%%%%%%%%%%%%%%%%%%
%%%%%%%%%%%%%%%%%%%%%%%%%%%%%%%%
%%%%%%%%%%%%%%%%%%%%%%%%%%%%%%%%

We consider now  films with nominal carrier concentration patterns consisting on various zones in the overdoped (OD) range.

Similarly as in the previous Section, we ran simulations  scanning numerous different patterns and selected the ones with best bolometric performance. We detail in \Table{table} the geometry parameters defining them via \Eqs{pj} and\NoEq{lambdap}. Their $\avp(x)$ profiles  are displayed in  the first column of \Fig{fig:OD}.

The first pattern, that we label  \ODsixzone, is composed by six zones that produce the same operating temperature range  $T^+-4.2\,{\rm K}$ than its underdoped counterpart (about 8\,K). The \Fig{fig:OD}(f) displays the computed $\Rsq(T)$ curve, \Fig{fig:OD}(g) the resulting output versus incident power for CCM  operation and \Fig{fig:OD}(h) the  output versus incident power for the CVM case. As in the underdoped case, already this relatively simple 6-zone option significantly improves the bolometric performance with respect to the non-patterned film (\ODonezone), mainly for \Pmin, \Pmax\ and DR (the latter improving by about one order of magnitude). The patterning even improves the TCR sensibility.

To complete our study, we also considered 10-zone patterns. Our computations resulted in the structure \ODtenzone\ (see  \Table{table} and third row of \Fig{fig:OD}). It may be seen that with respect to \ODsixzone\ this 10-zone pattern again improves \Pmax\ and DR, though at the expense of a slight increase of \Pmin. This structuration has the largest $T^+$ of all the reported in this paper, $T^+-4.2\,{\rm K}=22\,{\rm K}$. We conclude that, also in the overdoped case, 10-zone patterns have the potential of moderate improvements over the 6-zone option, to be pondered against the probable increase in manufacturing difficulty.

Similarly as for the underdoped case, our patterns allow operation in a simple CCM mode and in an ohmic region of the material, with no obvious advantage when using the more elaborate CVM. The reason for this may be traced back to  the operating range being wide in temperature, what makes  less critical the self-heating effects when compared to TES sensors with much narrow transitions. Naturally this wide operating range is also key for the large obtained DR, competitive with some of the best until now reported for TES bolometers\cite{fukuda} without apparent need for SQUID readout stages.

Some important differences with respect to the underdoped case are also observed: For instance, the \Rsq\  values in the upper part of the transition are smaller by about one order of magnitude, and also have different $T$-dependence after the linear transition region. (This in fact is one of the main reasons  for the values of patterning parameters $\gamma$ and $\beta$ being quite different to the underdoped case; the other being the \avp=1/8 depression of \Tc). The variation in \Rsq\ values are also  responsible for the quite different optimal bias voltages for the CVM modes in the underdoped and overdoped cases, a fact that may be of relevance in applications. Also, the  \Pmin\ of the underdoped structurations significantly differ from those of their overdoped counterparts, especially for CVM operation.

\section{Conclusions and additional  considerations for future work \label{Sec:Conclusions}}

In summary, we have presented methods to  calculate the effects of carrier density  structuration and patterning over the resistive transition \RT\ of thin films of the high-temperature superconductor \LSCOf\ (\LSCO) and their corresponding characteristics as sensing surface for resistive transition-edge bolometer detectors (TES) operating at a base temperature 4.2\,K  convenient for liquid-He cryogenics.  We also deploy ad-hoc computer search algorithms to discover the carrier density maps that optimize such sensing performance. We consider  two major contributions to  structuration: The Gaussian-random nanoscale disorder associated to carrier density nonstoichiometry, plus  the imposition of regular arrangements of zones with different nominal carrier densities (patterning). 

When applied to non-patterned films, our model is in good agreement with existing  measurements of \RT. 

When considering patterns composed of various zones of nominal carrier densities, our methods identify designs with either underdoped or overdoped carrier densities that markedly improve the saturation power and dynamic range of the TES detector and present  favorable sensitivity even in the easy-to-implement constant current mode of measurement.

Our results indicate that  structured \LSCO\ thin films are promising candidates  for TES operating at liquid-He temperature and, therefore, we suggest experimental workers to pursue their fabrication using our  pattern designs as guidelines.  As such, we believe that various considerations for those further works may be pertinent here:  {\it i)\/} Firstly, note that, as  already  mentioned in this article,   for a good  account of the resistive transition  it is crucial to take random disorder into account. We showed that our  procedures (here including in particular our \Eq{dp} for estimating the intrinsic, unavoidable disorder  due to nonstoichiometry for each nominal \avp) are accurate for  good quality films with disorder in the intrinsic limit (see Sect.~\ref{Sec:OneZone} or \Fig{fig:test}).  However, it is conceivable that maybe some pattern fabrication methods could, as a side-effect, induce  additional disorder contributions. Thus, it seems prudent that experimentalists check for possible nonuniversal additional contributions to \Eq{dp} in their films (for what we suggest to measure the resistive transition after inducing just a uniform 1-zone pattern). {\it ii)\/} Secondly, experimental details such as sample thickness may also induce small changes for both $\Tc(p)$ and $\rhon(T,p)$ with respect to our \Eqs{Tcp} to\NoEq{bkgdos}. We do not find however strong changes in our results when considering moderate changes for the parameters in those equations (those expected for moderate deviations from thickness 100~nm) unless that the \avpr\ values must be shifted to match the operational temperature 4.2\,K (in other words, for those small changes the newly obtained \avpr\ patterns basically shift to reproduce the former $\Tc(\vec{r})$ pattern).

In any case, the authors hereby express their willingness to help, if contacted, in repeating the type of simulations detailed in this article with parameters or phenomenological relationships custom-tuned to the possible specificities of those eventual experimental efforts.

\mbox{}\\
\mbox{}{\large\bf Acknowledgements.-- } {We acknowledge support by University of Santiago de Compostela, Project 2024-PU036 `Propiedades de materiais supercondutores micro e nanoestruturados'.  MMB was supported by Ministerio de Universidades, Spain, through National Program FPU (grant  number FPU19/05266).}

\end{document}